# Droplet charging in stratiform clouds

K.A. Nicoll[1,2] and R.G. Harrison[1]

1. Department of Meteorology, University of Reading, Reading, UK
2. Department of Electronic and Electrical Engineering, University of Bath, Bath, UK

**ABSTRACT:**
The role of droplet charge in stratiform clouds is one of the least well understood areas in cloud microphysics and is thought to affect cloud radiative and precipitation processes. Layer clouds cover a large proportion of the Earth's surface and are important in regulating the planetary radiation budget. Using a new remote sensing method developed at our University Observatory, we demonstrate that charge in the base of stratiform clouds is typically of negative polarity, as expected from theory considering the vertical current flow into and out of the cloud.

More detailed vertical charge structure of layer clouds can be found using balloon-carried instruments. Our previous research using in situ balloon observations has demonstrated that, on average, the bulk charge polarity and location agrees with theoretical predictions of positive charge at the upper edge and negative charge at the lower edge. Here we present optical and charge measurements of droplets from a variety of stratiform clouds, demonstrating the typical variability which is observed.

Although specially instrumented balloons are effective at determining stratiform cloud charge, they only provide a snapshot of the charge at any one time. In order to fully evaluate charge effects on cloud microphysics an extended series of surface and airborne measurements is required, which the combination of our new remote sensing technique and airborne instrumentation can now provide.

## INTRODUCTION

Stratiform clouds play a significant role in regulating Earth's climate as they are the most abundant cloud type, typically covering 30% of the globe (e.g. Klein and Hartmann, 1993). An often neglected property of stratiform clouds is that they are electrically charged, typically at their upper and lower horizontal boundaries, which results from vertical current flow in the global atmospheric electric circuit. As the conduction current, $J_c$, flows through the electrical conductivity transition between clear and cloudy air it generates layers of space charge at the conductivity gradients according to Gauss' law (e.g. Zhou and Tinsley, 2007). Assuming steady-state electrostatically, no vertical motion within the cloud and no horizontal divergence of the current density $J_c$, the charge per unit volume $\rho$ is

$$\rho = -\varepsilon_0 J_c \frac{d}{dz}\left(\frac{1}{\sigma_t}\right) = \varepsilon_0 J_c \left(\frac{1}{\sigma_t^2}\right)\frac{d\sigma_t}{dz} \qquad (1),$$

where the total conductivity is $\sigma_t$, $z$ is height, and $\varepsilon_0$ is the permittivity of free space. Equation (1) demonstrates that the cloud edge charge is negative at cloud base and positive at cloud top (assuming that $z$ is positive upwards). According to the simple electrostatic theory in equation (1) the cloud edge charge depends on the vertical conduction current density, the local ionization rate (which controls the conductivity) and the local meteorological conditions controlling the cloud properties (in particular the sharpness of the clear air-cloud boundary, given by $d\sigma_t/dz$). The cloud droplet charge resulting is expected to affect cloud microphysics through droplet coalescence (Harrison et al, 2015) and droplet-aerosol scavenging (Tinsley et al, 2000), which can ultimately affect cloud properties such as cloud height, cloud lifetime and even precipitation processes.

In order to more fully understand the effects of charge on cloud microphysical and, ultimately, any effect on climate processes, more observations of stratiform cloud edge charging are required. This has typically been achieved sporadically by airborne measurements from balloon (e.g. Jones et al, 1959; Nicoll and Harrison, 2010; Nicoll and Harrison, 2016) and aircraft (Imyanitov and Chubarina, 1967), however a new method of remote sensing of cloud base charge (Harrison et al, 2017) allows quantification of cloud edge charging in a more continuous manner. This paper will assess the validity of the remote sensing technique for cloud base charge by examining several case studies of balloon flights through a long lived layer of stratiform cloud. It will also demonstrate that cloud edge charge is highly sensitive to cloud structure, and differs even between clouds which appear visually and meteorologically similar.

**REMOTE DETECTION OF CLOUD BASE CHARGE**

Charge in the base of stratiform clouds can be detected remotely by exploiting the influence that the charge has on the near surface vertical atmospheric electric field $E_z$. To follow the usual convention of fair weather atmospheric electricity, the electric field will be considered as the Potential Gradient, PG ($-E_z$) which is positive in fair weather. Harrison et al (2017) demonstrated that cloud base charge can be determined using a combination of PG measurements (typically measured by an electric field mill) and a laser based ceilometer to measure the height of the cloud base. The ceilometer retrieves the amount of backscattered power at different time intervals after a pulse of laser light is emitted upwards, allowing a vertical profile of the atmosphere's properties beneath a cloud to be imaged: the greatest backscatter is generated by the cloud droplets, which is used to determine the position of the cloud edge and the cloud base height. An example of the backscatter profile retrieved from a ceilometer under stratiform cloud conditions is given in Figure 1(a). For extensive and persistent layer clouds having cloud base heights below about 1 km, the PG and cloud base height can be well correlated (Figure 1(b) and (c)). This is because the changes in height occur more slowly than the timescale for the cloud base charge to change, and therefore the electrostatic effect is similar to that of a fixed charge being moved towards and away from the field sensor.

The relationship between the PG and ceilometer cloud base height shown in Figure 1 (b) and (c) can be understood in terms of the induced effect of the steady cloud base charge, *i.e.,* as the cloud base lowers, the surface PG will be suppressed by the negative charge in the cloud base. Conversely, as the cloud base rises, the suppressing effect on the PG is reduced, until the cloud base charge is sufficiently distant for no effect to be observable and the PG returns to its undisturbed fair weather value.

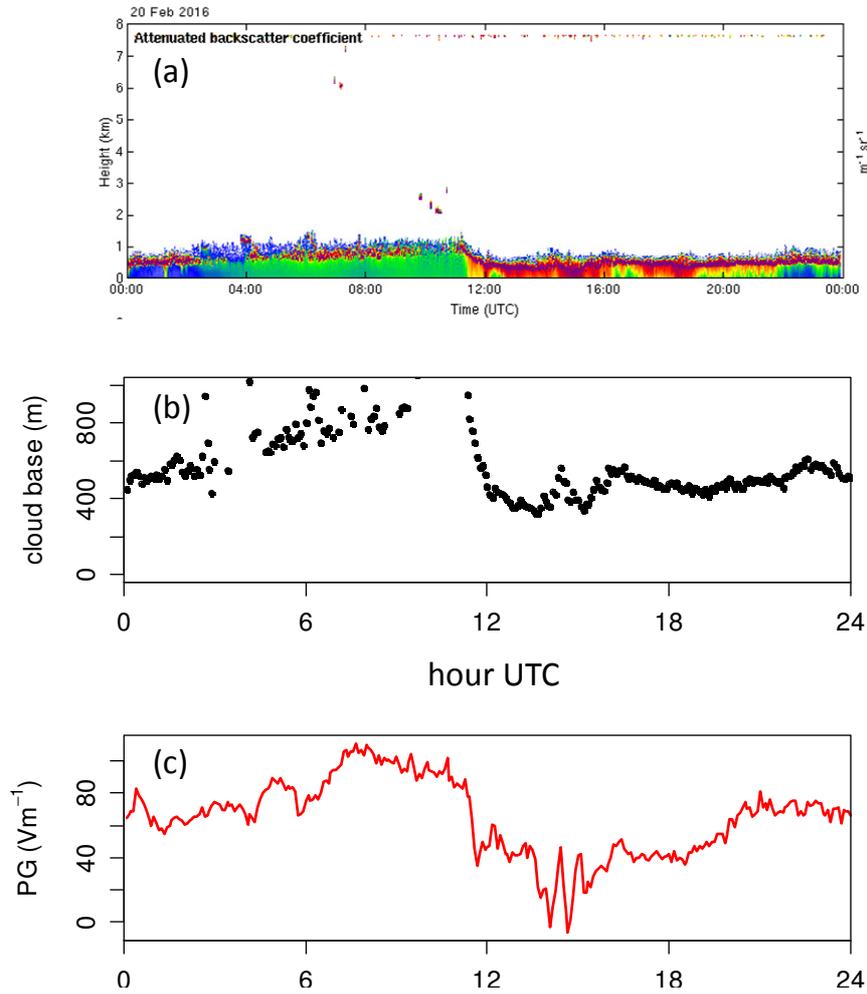

Fig. 1. Time series (plotted against hour of day) from 20th February 2016 at Reading, UK during persistent cloud with cloud base below 1000 m.   (a) Backscatter retrieved using a CL31 ceilometer, (b) cloud base height derived by the ceilometer algorithm from the backscatter and (c) Potential Gradient (PG) measured at the surface using a JCI131 field mill.   (The cloud base height and PG were recorded as 5 minute average values.)

The response of the surface PG to the cloud base charge can be modelled in the simplest case by assuming a single fixed charge of some appropriate geometry in the cloud base, e.g. a disk of charge. By applying the charged disk model to one year of PG and cloud base height data, Harrison et al (2017) demonstrated that the majority of the base of stratiform clouds below 1000m at Reading, UK, are negatively charged (in agreement with equation (1)), with a median cloud base charge per unit area of -0.86 nC m$^{-2}$.   Although this method can remotely detect cloud base charge, retrieved charge density requires an assumption about the depth of the charged layer to be made (which can only be determined from in-situ measurements) in order to estimate the space charge per unit volume (i.e. space charge density).   In order to verify the validity of this method for determining cloud base charge density a comparison with in-situ charge measurements is therefore required.

**DIRECT OBSERVATIONS OF CLOUD BASE CHARGE**

Balloon-borne instrumentation provides a route to in-situ measurements of cloud edge charge and does not have the limitations of the remote sensing method, which is limited to clouds with cloud base below 1000m and responds to the dominant lower cloud charge. However, the sporadic sampling nature of balloon measurements means that continuous observations of cloud charge are not generally possible by this method. The good vertical resolution (~5 m) which balloon measurements can provide are, however, very well suited to the purpose of better understanding the processes which control stratiform cloud edge charging. This has been demonstrated by recent balloon measurements which have established the presence of layer cloud edge charge at three locations in both hemispheres (Nicoll and Harrison, 2016) and shown that, on average, positive charge (+32pCm$^{-3}$) is found at cloud top and negative at cloud base (+24pCm$^{-3}$) in agreement with the electrostatic theory in Equation 1. The average space charge profile obtained from fifteen stratiform clouds from this work is shown in Figure 2. Nicoll and Harrison (2016) also demonstrated an asymmetry between charge at cloud top and cloud base (with cloud tops generally being more highly charged than cloud bases), thought to be associated primarily with cloud tops generally being more well defined meteorologically (e.g. by temperature inversions, and therefore possessing larger gradients in conductivity) than cloud bases.

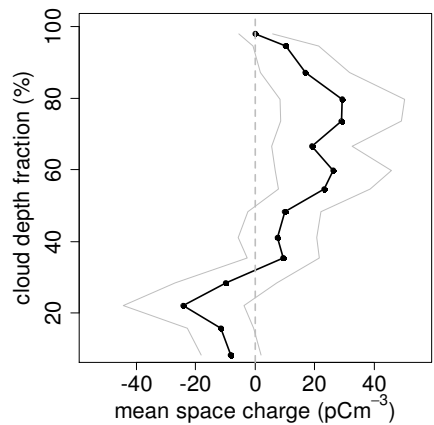

Figure 2. Profiles from 15 cloud and charge sensor balloon flights through stratocumulus clouds with height <3km, adapted from Nicoll and Harrison (2016). The vertical axis denotes height normalised by cloud depth, found by dividing each cloud layer into 15 evenly spaced altitude layers, where 0% denotes cloud base and 100% cloud top. Grey solid lines show two standard errors on the mean values.

**ANTARCTIC CLOUD CHARGE MEASUREMENTS**

A recent field campaign to Halley, Antarctica provided a rare opportunity to perform simultaneous measurements of PG, ceilometer cloud base height and in-situ cloud charge measurements from specially instrumented balloon flights through a layer of low level stratiform cloud which lasted several days. Halley is a remote site run by the British Antarctic Survey, located on the Brunt Ice Shelf (75.58°S, 26.66°W) with snow cover all year round. PG measurements were made there with a JCI131 electric field mill (which was deployed from Feb 2015-Jan 2017) and a Vaisala CT25K ceilometer measured backscatter and cloud base height. In-situ cloud measurements were made from a number of specially instrumented Vaisala RS92 radiosonde balloons. These carried optical cloud droplet sensors (Harrison and Nicoll, 2014) to determine the cloud droplet microphysical properties, as well as miniaturized space charge sensors (Nicoll 2013) to measure the in-cloud space charge density. These sensors were interfaced to standard RS92 meteorological radiosondes using the PANDORA data acquisition system developed at the University of Reading (Harrison et al, 2012) and flown beneath a 200g latex balloon.

Figure 3 shows a time series of cloud base height measurements from the ceilometer on 20$^{th}$ and 21$^{st}$ Feb 2015 at Halley (year day 51 and 52) demonstrating the presence of a long lived layer of stratiform cloud which lasted two days. The height of the cloud base varies very little (of order 200m) between the two days. Also shown are the PG measurements from the JCI131 electric field mill, which shows variability throughout the period of cloud. It should be noted that the wind speed was less than 10m/s and the visual range greater than 40km at all times during this period, minimizing the potential for blowing snow to affect the PG measurements.

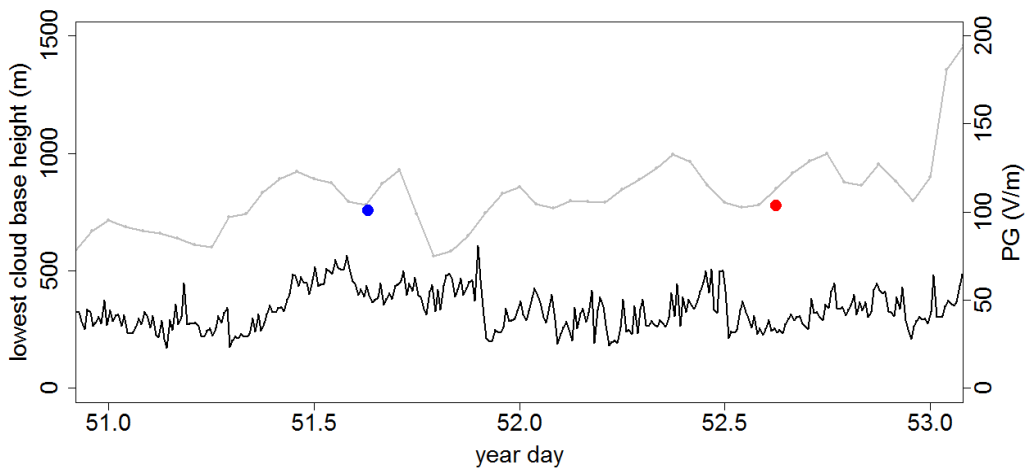

Figure 3. Measurements from Halley, Antarctica during 20$^{th}$ and 21$^{st}$ Feb 2015 (day 51 and 52 respectively) during a long lived layer of stratiform cloud. Cloud base height measured from the Vaisala CT25K is shown in grey and PG measured by a JCI131 electric field mill in black. Blue and red dots show the height of the cloud base determined from the optical cloud sensor on board specially instrumented balloons, plotted at the time at which the balloon passed through the cloud layer.

By applying the charged disk model of Harrison et al (2017) to the cloud base height and PG data from Halley during 20$^{th}$ and 21$^{st}$ February, it is possible to derive an estimate of the mean cloud base charge for each of the days in question using the remote sensing method. The modelled fit to the data gives a mean cloud base charge of -9.93nCm$^{-2}$ derived for 20$^{th}$ Feb and -14.59nCm$^{-2}$ derived for 21$^{st}$ Feb.

To provide in-situ measurements of cloud edge charge to compare the derived remote sensing values to, the specially instrumented radiosondes were launched at 15:20 on 20$^{th}$ and 15:01 on 21$^{st}$ February (blue and red dots in Figure 3 respectively). Vertical profiles of the thermodynamic properties of the cloud layer are shown in Figure 4. It is seen that even though these measurements were made ~24 hours apart, the temperature and relative humidity (RH) profiles through the cloud layer are almost identical, suggesting very little change to the overall thermodynamic structure of the cloud.

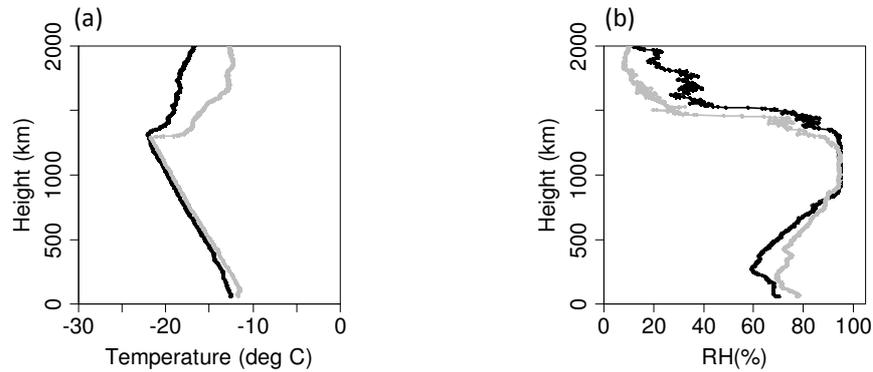

Figure 4. Vertical profiles of thermodynamic quantities from two balloon flights through a persistent layer of stratiform cloud at Halley.   Grey = 15:20 UT 20th Feb 2015 (day 51) and black = 15:01 UT 21st Feb 2015 (day 52).

The cloud droplet and electrical measurements from the two balloon flights are shown in Figure 5. Figure 5 (a) and (c) demonstrate that the cloud layer was well defined optically on both days, with very clear cloud-air transitions (and therefore conductivity gradients) at the cloud edge boundaries. Figures 5 (b) and (d) (black lines) show the in-situ cloud charge measurements which, unlike the optical cloud droplet profiles, show very different behavior between the two days.  On 20th Feb (Figure 5(b)) the in-cloud space charge density ranges between ~ -30 to +30 pCm$^{-3}$, with little structure around the upper and lower horizontal cloud edge boundaries; conversely Figure 5(d) demonstrates a well defined negative layer of charge in the cloud base (up to -110 pCm$^{-3}$) on 21st Feb, which is coincident with the cloud edge transition zone as shown by the optical droplet sensor.   Towards the cloud top, the charge is predominantly positive, although smaller in magnitude than the cloud base.

The grey lines in Figure 5(b) and (d) show the predicted space charge calculated using the ion balance equation to estimate conductivity, and equation (1) (see Nicoll and Harrison, 2016 for details using this method).   Cloud droplet number concentration is derived from the optical cloud droplet sensor backscatter, assuming a droplet radius of 10μm, and the ionization rate is estimated to be 2 ion pairs cm$^{-3}$ s$^{-1}$, with a background aerosol concentration of 1000 cm$^{-3}$.   Figure 5(d) demonstrates that the location of the predicted and measured space charge is similar, and that the asymmetry in the predicted space charge (37 pCm$^{-3}$ at cloud top and -48 pCm$^{-3}$ at cloud base) matches well with the increased charge in the cloud base compared to cloud top.   This is attributed to the sharper conductivity gradient at cloud base (as observed from figure 5 (c)) than cloud top.   The difference in the magnitude of predicted and observed charge arises from the assumptions necessary in the calculations.

To perform a direct comparison with the cloud base charge detected by the remote sensing method the space charge per unit area is required (in Cm$^{-2}$), rather than the space charge per unit volume (Cm$^{-3}$), which the balloon charge sensor measures. To derive the in-situ space charge per unit area the median space charge in the cloud base is calculated (using only the absolute values of charge) and therefore multiplied by the depth of the cloud base. The depth of the cloud base is defined as the depth over which the negative region of charge was predicted by theory (i.e. from the grey line in Figure 5(b) and (d) - this is 234m for 20th Feb and 168m for 21st Feb).    A comparison of the remotely sensed values of cloud base charge with the in-situ balloon measurement charge is given in Table 1.

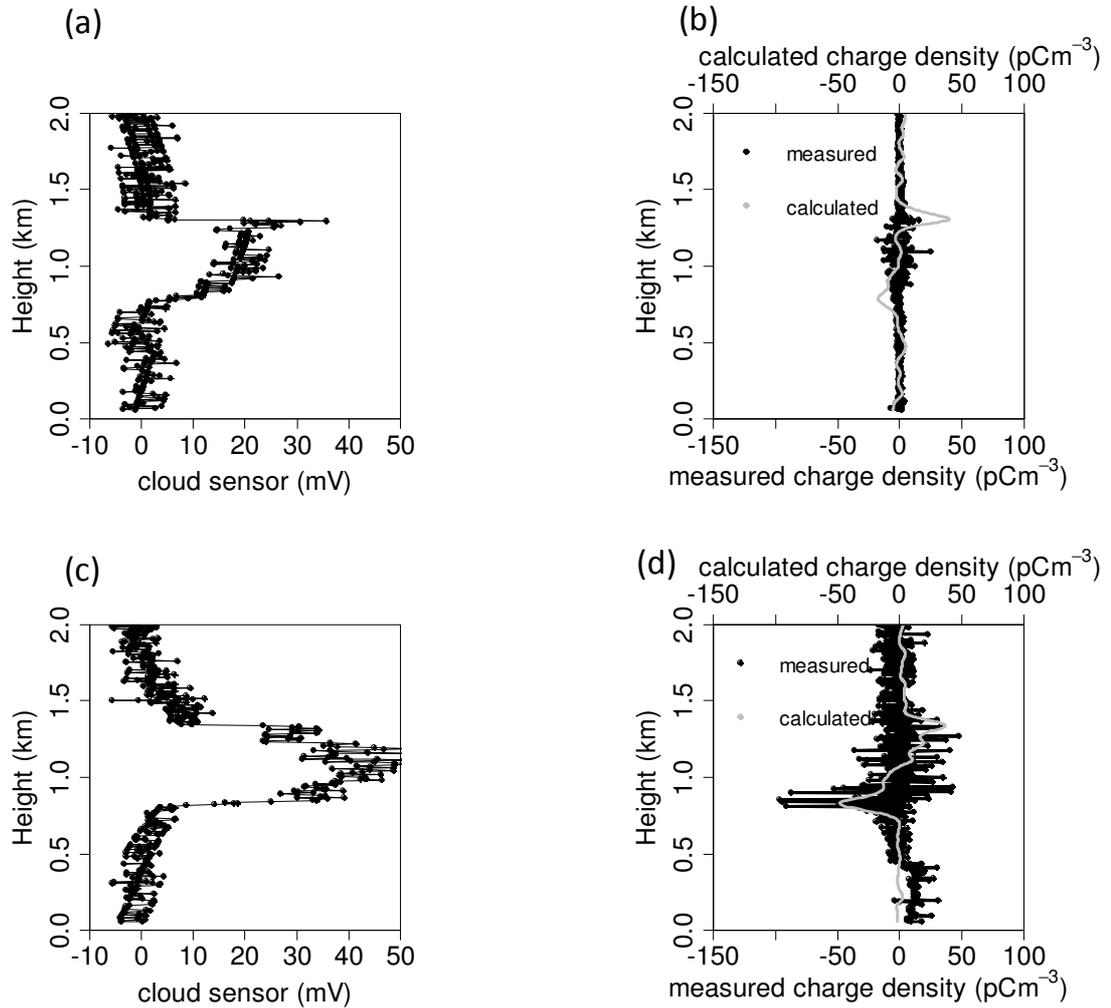

Figure 5. Vertical profiles through a long-lived stratiform cloud at Halley on 20$^{th}$ ((a) and (b)) and 21$^{st}$ ((c) and (d)) February 2015. (a) and (c) show the backscatter voltage from the optical cloud sensor. (b) and (d) show the measured space charge density (black) alongside the hypothetical calculated space charge derived from the optical cloud sensor profiles and conductivity considerations using the ion-balance equation (in grey).

| Date | Remote method cloud base charge | In-situ method cloud base charge |
|---|---|---|
| 20$^{th}$ Feb 2015 | -9.9 nCm$^{-2}$ | -0.36 nCm$^{-2}$ |
| 21$^{st}$ Feb 2015 | -14.6 nCm$^{-2}$ | -3.6 nCm$^{-2}$ |

Table 1. Comparison of cloud base charge derived remotely from PG and ceilometer measurements using the method of Harrison et al (2017); and measured in-situ from balloon borne charge sensors.

## DISCUSSION

The comparison between the remotely sensed and in-situ measured cloud base charge measurements demonstrates that in general the remote sensing method yields greater amounts of charge, but that these are still within an order of magnitude of the directly measured values. The general trend, however, is correct in that both methods agreed that the cloud base on 21$^{st}$ Feb was considerably more charged than on 20$^{th}$ Feb. It is expected that there will be some discrepancy between the two methods due to the number of assumptions made, particularly in the case of the remote sensing technique (e.g. that the cloud charge resembles a charged disk). There is also the challenge of what is the appropriate depth of the cloud edge charge region to use for the in-situ case, which has many potential definitions.

This work also demonstrates the differences in cloud edge charge that can exist within clouds which appear similar, both visually and thermodynamically. The likely explanation for such differences is related to the cloud droplet profiles as well as the turbulent characteristics of the cloud. For example, close inspection (not shown here) of the base of the cloud droplet profile on 20$^{th}$ suggests a profile which is indicative of droplet growth and therefore that updrafts are operating within the cloud at this time. This would act to mix charge in the cloud base, inhibiting the formation of a well-defined layer of negative charge at the cloud base. Conversely, the optical cloud sensor profile on 21$^{st}$ suggests no growth of droplets and therefore very little mixing within the cloud, enforcing the well-defined negative charge layer. The predominantly positive charge throughout the top and middle of the cloud suggests that gravitational settling of positive charge from the top of the cloud may also be active in distributing positive charge down throughout the cloud.

## CONCLUSIONS

This paper has demonstrated a direct comparison of cloud base charge derived from a new remote sensing technique with directly measured in-situ charge measurements during a long lived cloud layer in Antarctica. Results show that the two methods yield values within the correct order of magnitude of each other, demonstrating the potential use of the remote sensing method for near continuous monitoring of cloud base charge in low level stratiform clouds.

The measurements reported here also demonstrate the variability in cloud charge apparent within stratiform cloud layers which appear very similar from the outset (both visually and thermodynamically), and that the details of the cloud droplet profiles and meteorological mixing processes within such clouds are key to controlling the cloud charge profile. Such processes must therefore be considered when assessing the role of external influences on cloud edge charging (e.g. through $J_c$ and conductivity changes).

## ACKNOWLEDGMENTS

K.A.N. acknowledges Natural Environment Research Council (NERC) support through an Independent Research Fellowship (NE/L011514/1 and NE/L011514/2) and the assistance of colleagues at the British Antarctic Survey (including David Maxwell, Steve Colwell and Mervyn Freeman) for assistance with data, logistics and balloon launching at Halley, funded through a NERC Collaborative Gearing Scheme (CGS) grant.